\documentclass[journal,10pt]{IEEEtran}
\usepackage[nosumlimits]{amsmath}
\usepackage{amssymb}
\usepackage{bbm}
\usepackage{bm}
\usepackage[capitalize]{cleveref}

\DeclareMathOperator*{\argmax}{arg\,max}
\newcommand{\E}{\mathbb{E}}
\usepackage{pgfplots}
\usepackage{algorithm}
\usepackage{algpseudocode}

\begin{document}
\title{Minimizing the Age of Information from Sensors with Common Observations}

\author{Anders~E.~Kal{\o}r,~\IEEEmembership{Student Member,~IEEE,} and Petar~Popovski,~\IEEEmembership{Fellow,~IEEE}%
\thanks{The authors are with the Department of Electronic Systems, Aalborg University, Denmark. (e-mail: aek@es.aau.dk; petarp@es.aau.dk)}%
\thanks{The work has been in part supported by the Danish Council for Independent Research, Grant Nr. 8022-00284B SEMIOTIC and
by TACTILENet (Grant no. 690893), within the Horizon 2020 program.}}

\maketitle

\begin{abstract}
  We study the average Age of Information (AoI) in a system where physical sources produce independent discrete-time updates that are each observed by several sensors. We devise a model that is simple, but still capable to capture the main tradeoffs. Two sensor scheduling policies are proposed to minimize the AoI of the sources; one in which the system parameters are assumed known, and one in which they are learned. Both policies are able to exploit the common sensor information to reduce the AoI, resulting in large reductions in AoI compared to common schedules.
\end{abstract}
 

\section{Introduction}
\begin{figure}[tb]
  \centering
  \includegraphics{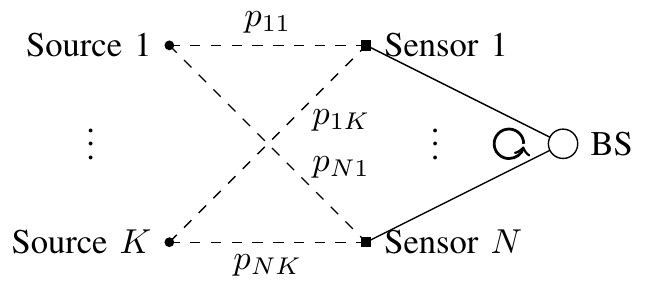}
  \caption{System model with $K$ sources and $N$ sensors.}\label{fig:sysmodel}
\end{figure}

The rising number of use cases for Internet-of-Things (IoT) and Machine-Type Communication (MTC) has given rise
to a metric termed Age of Information (AoI). It  
describes the timeliness of information at
the receiver, see
e.g.~\cite{real-time-status-how-often,updating-iot-correlated-sources}. Common to most AoI scenarios is that the information should be delivered in a timely manner, and usually only the most recent information is of interest.
In this letter, we consider the scenario depicted in~\cref{fig:sysmodel}, where $K$ sources generate independent status updates at
random time instants. Updates from source $k$ are observed by
sensor $n$ according to a Bernoulli process with parameter $p_{nk}$. Each sensor stores the most
recently observed update from each source.
In each time slot, the Base Station (BS) can schedule one sensor
to transmit its most recent update from a specific source. If the
update is more recent than the most recent update known to the
BS, the AoI of the source is reduced.
Since each source update may be observed by multiple sensors, the updates stored by the sensors are correlated. For
example, in the case with two sources and two sensors, the most
recent update from Source 1 may be known to both Sensor 1 and 2, while
the most recent update from Source 2 may be known only to Sensor 2.
This is a simplification of a scenario that is applicable in many IoT systems, such as systems
where the sensors act as gateways/relays for the source updates, or
data collection systems based on concentrators.
Unlike the independent updates, the fact that source
updates are observed by multiple sensors introduces
redundancy in the network, which can be used to reduce the AoI.

The AoI metric~\cite{real-time-status-how-often}
has been used to study timely update policies in various multi-user contexts. In~\cite{real-time-multiple-sources} the sources communicate to a receiver through a shared 
queue, \cite{status-updates-multiaccess} considers random access system, and scheduled access is considered in~\cite{kadota2018optimizing} and~\cite{talak2018scheduling}.
Temporally correlated updates from a single source
are considered in~\cite{realtime-update-correlated-source},
and in~\cite{minimize-aoi-correlated-camera} the authors
consider a wireless camera network in which the cameras have
overlapping fields of view. Cameras with overlapping scenes are connected to the same destination nodes, which delay the processing until frames from all cameras that capture a specific scene have been received. A scenario where two sensors observe correlated Gaussian processes is presented in~\cite{updating-iot-correlated-sources}. The two sources sample
and transmit updates periodically at fixed (but possibly different)
intervals, and the authors study the estimation error resulting from various update strategies. The system in~\cite{updating-iot-correlated-sources} differs from our system model as we treat centralized scheduling with arbitrary scheduling intervals, as well as discrete updates rather than a continuous process.

\section{System Model}\label{sec:sysmodel}
We consider a system with $K$ sources, $N$ sensors and one base
station as depicted in \cref{fig:sysmodel}. The sources generate updates
according to independent Poisson processes with rates
$\lambda_1,\ldots,\lambda_K$. An update generated by source $k$ is
observed by sensor $n$ with probability $p_{nk}$.
Denote by $\Phi_{nk}=\{\phi_1^{(nk)},\phi_2^{(nk)},\ldots\}$ the set
of time instances at which sensor $n$ observes an update from source
$k$. We then define the AoI of source $k$ at
sensor $n$ at time $t$ as $\Delta_{nk}(t)=t-\max_i \phi_i^{(nk)}$.

In each time slot $t=1,2,\ldots$, the BS can schedule a sensor to
transmit its most recent update from a specific source. The transmission is error-free and instantaneous.
Let $u_{nk}(t)=1$ if sensor-source pair $(n,k)$ is
requested at time $t$, and $u_{nk}(t)=0$ otherwise. We constrain the transmissions to be orthogonal by setting $\sum_n\sum_k u_{nk}(t)\le 1$.
The AoI of source $k$ at the BS is then given by the
following process:
\begin{equation}\label{eq:aoiprocess}
  \Delta^{(k)}(t)=\min\left(\sum_{n} u_{nk}(t)\Delta_{nk}(t), \Delta^{(k)}(t-1)+1\right).
\end{equation}
To simplify the notation we let $\Delta^{(k)}(0)=0$ for
all $k$, and define the average AoI at the BS at time $t$ as
\begin{equation}
  \Delta'(t)= \frac{1}{K}\sum_{k=1}^K \Delta^{(k)}(t).
\end{equation}

\begin{figure}[tb]
  \centering
  \includegraphics{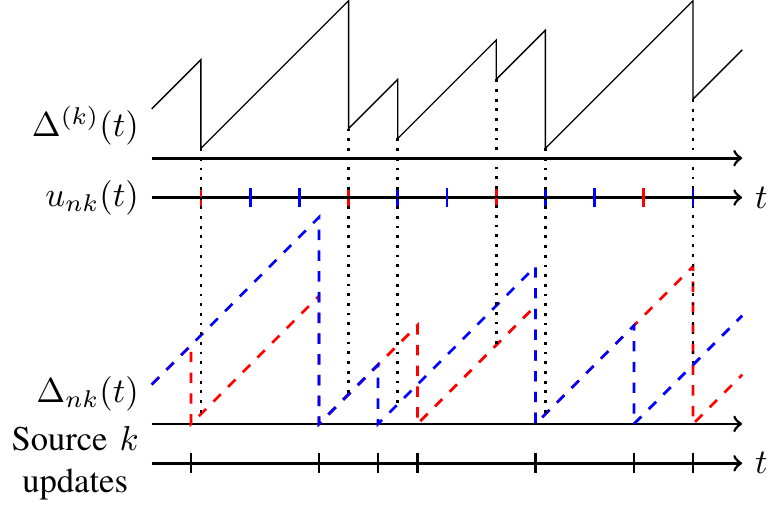}
  \caption{The AoI for a source $k$ observed by two sensors, red and
    blue. Each sensor is scheduled according to an (arbitrary) policy
    indicated by $u_{nk}(t)$.}\label{fig:aoi}
\end{figure}

A realization of the process in \cref{eq:aoiprocess} with a single source and two sensors is
illustrated in \cref{fig:aoi}, where AoI at each sensor $\Delta_{nk}(t)$ is illustrated by red and blue lines. When a sensor is scheduled, indicated in the $u_{nk}(t)$ axis, the AoI at the BS $\Delta^{(k)}(t)$ is updated according to \cref{eq:aoiprocess}.

We seek policies $\pi$ that characterize
the scheduling, $u_{nk}(t)$, and minimize the
long-term average of $\Delta'(t)$.
Mathematically, we aim at solving the problem
\begin{equation}\label{eq:overallproblem}
  \underset{\pi}{\text{minimize}} \lim_{T\to\infty} \frac{1}{T}\sum_{t=1}^T \E\left[\Delta'(t)\mid\Lambda(t-1)\right],
\end{equation}
where $\Lambda(t)$ denotes the system state at time $t$ containing the AoI at the BS and the \emph{sampling histories}, $\mathcal{S}_{nk}(t)$, i.e. $\Lambda(t)=\left(\mathcal{S}_{11}(t),\ldots,\mathcal{S}_{NK}(t), \Delta^{(1)}(t),\ldots,\Delta^{(K)}(t)\right)$. $\mathcal{S}_{nk}(t)$ will be defined later.

We first consider the optimal policy under the assumption that the
BS knows the parameters $\lambda_k$ and $p_{nk}$. We then
relax this requirement and consider a model-free policy inspired by
contextual bandits that learns the parameters by
exploring the system while keeping the AoI low.

\section{Policies with Known Parameters}\label{sec:knownparams}
We first consider the set of policies for which the BS is assumed to know the system parameters $\lambda_k$ and $p_{nk}$.
In order to minimize the average AoI in \cref{eq:overallproblem},
observe that, since the updates are memoryless and independent accross sources, the optimal policy is to maximize the expected reduction in $\Delta'(t)$.
Let $(n(t),k(t))$ denote the sensor-source pair that is scheduled in time slot $t$, i.e.
$u_{n(t)k(t)}=1$. The optimal policy selects
\begin{align}\label{eq:optimalpolicystart}
  &(n(t),k(t))= \nonumber\\
  &\argmax_{n,k}\left\{\Delta^{(k)}(t-1)+1-E\big[\Delta^{(k)}(t)\mid\Lambda(t-1)\big]\right\},
\end{align}
where $u_{n(t)k(t)}$ is included in the expression for $\Delta^{(k)}(t)$ as in \cref{eq:aoiprocess}.
For simplicity, in the following we omit the dependency on $t-1$ in the expression for $\Lambda(t-1)$. Denote by $Z_{nk}$ the event $\Delta_{nk}(t)<
\Delta^{(k)}(t-1)+1$, i.e. that sensor-source pair $(n,k)$ has a more recent
observation. Using the law of total expectation and \cref{eq:aoiprocess} we have
\begin{align}
  (n(t),k(t))&=\argmax_{n,k} \Delta^{(k)}(t-1)+1\nonumber\\
  &\qquad-\E[\Delta_{nk}(t)\mid Z_{nk},\Lambda]\Pr(Z_{nk}\mid \Lambda)\nonumber\\
  &\qquad -
  \left(\Delta^{(k)}(t-1)+1\right)\left(1-\Pr(Z_{nk}\mid \Lambda)\right)\nonumber\\
  &=\argmax_{n,k}
  \Pr(Z_{nk}\mid \Lambda)\Big(\Delta^{(k)}(t-1)+1\nonumber\\
    &\qquad-\E[\Delta_{nk}(t)\mid Z_{nk},\Lambda]\Big).\label{eq:optimal1}
\end{align}
To obtain an expression for $\Pr(Z_{nk}\mid \Lambda)$, observe that due to the memoryless property of the Poisson updates, $Z_{nk}$ only depends on the sampling history since the previous reduction in $\Delta^{(k)}(t)$, or the last time sensor $n$ was scheduled if this is more recent. Denoting by $\tau_{nk}$ the time since
sensor-source pair $(n,k)$ was last scheduled, and letting
$\sigma_{nk}=\min(\tau_{nk},\Delta^{(k)}(t-1)+1)$, we define
\begin{equation}
  \mathcal{S}_{nk}=\{\tau_{mk}\mid m=1,2\ldots N\wedge m\neq n\wedge \tau_{mk}<\sigma_{nk}\}.
\end{equation}

\begin{figure}[tb]
  \centering
  \includegraphics{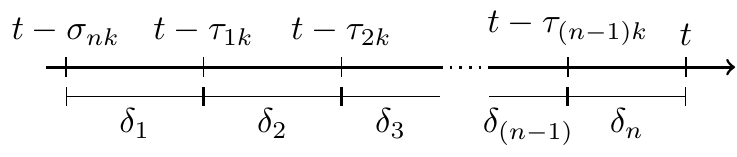}
  \caption{Notation used to derive the optimal policy.}\label{fig:optimal_policy_proof}
\end{figure}

Since both $\mathcal{S}_{nk}$ and $\Delta^{(k)}$ are independent across sources, we can
consider each source independently in the derivation of $\Pr(Z_{nk}\mid \Lambda)$. Without loss of
generality, suppose that the BS observes a new update from source $k$ at
time $t$ by sampling sensor $n$ after having unsuccessfully sampled sensor-source pairs
$(1,k),(2,k),\ldots,(n-1,k)$. Notice that we may sample other
sources between the samples of source $k$, and hence denote the time between
sampling sensor-source pairs $(j,k)$ and $(j+1,k)$ by $\delta_{j+1}$ as illustrated in
\cref{fig:optimal_policy_proof}. From the memoryless property of the
exponential distribution we have
\begin{align}\label{eq:prz1}
  \Pr(Z_{nk}\mid \Lambda) &= 1-\prod_{j=1}^n(1-\Pr(Z_{nk}^{(j)}\mid \Lambda)),
\end{align}
where $Z_{nk}^{(j)}$ is the event that sensor $n$ observed an update
from source $k$ during $\delta_j$.
Since per definition all samples up to $t$ were
unsuccessful, sensors $j,\ldots,n-1$ did not observe an update in
$\delta_j$. Denoting by $L_j$ the random number of source updates in
$\delta_j$ and using the fact that the sensor observations are
independent conditioned on $L_j$, it follows that the probability that
sensor $n$ also did \emph{not} observe an update in $\delta_j$ is given by
\begin{align*}
  1-\Pr(Z_{nk}^{(j)}\mid \Lambda)%
  &=\sum_{l=0}^\infty \frac{(\delta_j\lambda_k)^l e^{-\delta_j \lambda_k}}{l!}(1-p_{nk})^l\\
  &=e^{-\delta_j\lambda_k p_{nk}},
\end{align*}
where we used the definition of conditional probability and the fact
that $L_j$ is Poisson distributed. Inserting into \eqref{eq:prz1} yields
\begin{align}\label{eq:prz}
  \Pr(Z_{nk}\mid \Lambda) &= 1-\prod_{j=1}^n e^{-\delta_j\lambda_k p_{nk}}\nonumber\\
  &= 1-e^{-\lambda_k p_{nk} \sum_{j=1}^n \delta_j}\nonumber\\
  &= 1-e^{-\lambda_k p_{nk}\sigma_{nk}}.
\end{align}
A consequence of this expression is that $\sigma_{nk}$ completely
describes the sampling history of sensor-source pair $(n,k)$. In other
words, the fact that we have not observed anything from
previously scheduled sensors does not give us any information regarding the
likelihood that sensor $n$ has an update.

We derive $\E[\Delta_{nk}(t)\mid Z_{nk},\Lambda]$ in a similar fashion. By independence of the observed sensors and memoryless observations, the time since the last observation $\Delta_{nk}(t)$ follows an exponential distribution truncated to $[0,\sigma_{nk}]$:
\begin{align}\label{eq:edelta}
  \E[\Delta_{nk}(t)\mid Z_{nk},\Lambda] &= \frac{\int_0^{\sigma_{nk}} x \lambda_k p_{nk}e^{-\lambda_k p_{nk}x}\text{d}x}%
  {1-e^{-\lambda_k p_{nk}\sigma_{nk}}}\nonumber\\
  &=\frac{1-e^{-\lambda_k p_{nk}\sigma_{nk}}(\lambda_k p_{nk}\sigma_{nk}+1)}
  {\lambda_k p_{nk}\left(1-e^{-\lambda_k p_{nk}\sigma_{nk}}\right)}.
\end{align}
By substituting \labelcref{eq:prz} and \labelcref{eq:edelta} into
\cref{eq:optimalpolicystart} and rearranging we find that the optimal
policy is
\begin{align}\label{eq:optimalpolicy}
  (n(t),k(t))&=\argmax_{n,k} \left(\Delta^{(k)}(t-1)+1 - \frac{1}{\lambda_k
      p_{nk}}\right) \nonumber\\
  &\qquad \cdot \left(1-e^{-\lambda_k p_{nk}\sigma_{nk}}\right) +
  \sigma_{nk}e^{-\lambda_k p_{nk}\sigma_{nk}},
\end{align}
which can be solved by iterating over all $(n,k)$-pairs with $p_{nk}>0$.

\section{Policies with Unknown Parameters}\label{sec:unknownparams}
In this section,
we consider policies that do not require $\lambda_{k}$ and $p_{nk}$ to be known. To this end, we formulate the problem as a contextual bandit problem~\cite{sutton2018reinforcement}, in which the agent (the BS) observes context information (the state) prior to selecting an action, and
receives an immediate reward (the reduction in AoI) that is independent of previous actions. This is suitable for our problem as the AoI reductions are conditionally independent given the state and action.

Several policies have been proposed for solving contextual bandits,
ranging from the simple, sub-optimal
$\epsilon$-greedy~\cite{sutton2018reinforcement} to the optimal but more
complex Upper Confidence Bound (UCB)
algorithms~\cite{Li_linucb}.
In this work, to demonstrate the applicability of contextual bandits
to the problem, we restrict our focus on the simple $\epsilon$-greedy
policy, and remark that the performance may be improved by using
an algorithm with better performance guarantees.

Because the state space is continuous (and hence infinite), we
apply an approximate method that models the expected reward, $\E[R]$,
achieved by scheduling a specific sensor-source pair $(n,k)$ as a
parametric function. The agent then picks the sensor-source pair that yields the highest $\E[R]$. Inspired by \cref{eq:optimal1} we
model the expected reward as
\begin{align}
  \E[R\mid \Lambda,(n,k)]
  &\approx
  \hat{p}\left(\bm{\psi}_{(n,k)}^\mathrm{T}\bm{x}_{\Lambda,(n,k)}\right)\Big(\Delta^{(k)}(t-1)+1\nonumber\\
  &\qquad-\bm{\theta}_{(n,k)}^\mathrm{T}\bm{x}_{\Lambda,(n,k)}\Big)\nonumber\\
  &\triangleq \hat{r}\left(\bm{x}_{\Lambda,(n,k)},
  \bm{\theta}_{(n,k)}, \bm{\psi}_{(n,k)}\right),
\end{align}
where $\hat{p}(x)=1/(1+e^{-x})$ is the sigmoid function, $\bm{x}_{\Lambda,(n,k)}$ is a feature vector that
represents the current state and the action of scheduling pair $(n,k)$, and $\bm{\theta}_{(n,k)}$ and $\bm{\psi}_{(n,k)}$ are
unknown parameter vectors that need to be learned. $\hat{p}(\cdot)$
represents the probability of receiving a non-zero reward,
and $\bm{\theta}_{(n,k)}^\mathrm{T}\bm{x}_{\Lambda,(n,k)}$ is an estimate of the expected new age, conditioned on a non-zero reward, obtained by scheduling sensor-source pair $(n,k)$ in state $\Lambda$. While this linear model is unable to represent the
actual reward function in \cref{eq:optimalpolicy}, the goal is to
obtain an approximation that predicts which sensor-source pairs are
likely to yield a high reward.

The feature vectors
$\bm{x}_{\Lambda,(n,k)}$ are the input to the reward estimate $\hat{r}(\cdot)$ and represent the action of scheduling pair $(n,k)$ in the current state. Since $\sigma_{nk}$ is the only
state parameter that characterizes the probability and the age of a new observation (see \cref{eq:prz,eq:edelta}), we simply define
$\bm{x}_{\Lambda,(n,k)}=[1,\sigma_{nk}]^{\mathrm{T}}$, where the first entry represents the bias element that allows the function to be shifted from the origin.
While more complex features could be included, such as non-linear
transformations of $\sigma_{nk}$, such features have shown not to give
better performance in the scenario considered in \cref{sec:numres}.

\subsection{Training Algorithm}
In the $\epsilon$-greedy algorithm~\cite{sutton2018reinforcement},
shown in \cref{lst:algorithm}, the agent schedules with probability $1-\epsilon$ the sensor-source
pair that maximizes the expected reward, and with probability
$\epsilon$ a pair selected uniformly at random. Therefore, $\epsilon$ controls the
exploration/exploitation trade off and is usually set to a low value.
After performing an action and observing the resulting reward, $R$, the
agent updates the parameter vectors $\bm{\theta}_{(n,k)}$ and
$\bm{\psi}_{(n,k)}$ using stochastic gradient descent.
We update $\bm{\psi}_{(n,k)}$ using the cross-entropy loss function
with derivative $\frac{\partial}{\partial
  \psi_i}\ell(y,\bm{\psi}^{\mathrm{T}}\bm{x})=(\hat{p}(\bm{\psi}^{\mathrm{T}}\bm{x})-y)x_i$,
where $y$ is $1$ if $R>0$ and $0$ otherwise, i.e. $y=\mathbb{I}(R>0)$.
Similarly, $\bm{\theta}_{(n,k)}$ is updated with
the least-squares loss function with derivative $\frac{\partial}{\partial
  \theta_i}\ell(y,\bm{\theta}^{\mathrm{T}}\bm{x})=(\bm{\theta}^{\mathrm{T}}\bm{x}-y)x_i$
where $y=\Delta^{(k)}(t-1)+1-R$. Notice that $\bm{\theta}$ is only updated if the reward is non-zero.

\begin{algorithm}
  \caption{$\epsilon$-greedy algorithm.}\label{lst:algorithm}
\begin{algorithmic}
  \algblockdefx[WITHPR]{WithPr}{EndWithPr}[1]{\textbf{With probability} #1 \textbf{do}}{\textbf{End}}
  \algcblock{WITHPR}{Else}{EndWithPr}
  \algrenewcommand\algorithmicfor{\textbf{For}}
  \algrenewtext{EndFor}{\textbf{End for}}
  \algrenewcommand\algorithmicif{\textbf{If}}
  \algrenewtext{EndIf}{\textbf{End if}}
  \algrenewcommand\algorithmicelse{\textbf{Else}}
  \algrenewcommand\algorithmicwhile{\textbf{While}}
  \algrenewtext{EndWhile}{\textbf{End while}}
  \State \textbf{Input:} $\epsilon\ge 0$, step size $\alpha>0$, initial state $\Lambda$
  \State Initialize $\bm{\theta}_{(n,k)}\gets\bm{0}$,
  $\bm{\psi}_{(n,k)}\gets\bm{0}\quad \forall (n,k)$
  \For{$t=1,2,\ldots$}
  \WithPr{$\epsilon$}
  \State Draw $(n',k')$ uniformly at random
  \Else
  \State $(n',k')\gets \argmax_{(n,k)}
  \hat{r}\left(\bm{x}_{\Lambda,(n,k)},
  \bm{\theta}_{(n,k)}, \bm{\psi}_{(n,k)}\right)$
  \EndWithPr
  \State $\Delta^{(k)\prime}\gets \Delta^{(k)}(t)$
  \State Schedule $(n',k')$, observe reward $R$ and new state $\Lambda'$
  \If{$R>0$}
  \State $\hat{\Delta}_{nk}\gets \bm{\theta}_{(n',k')}^{\mathrm{T}}\bm{x}_{\Lambda,(n',k')}$
  \State $\bm{\theta}_{(n',k')}\gets \bm{\theta}_{(n',k')} - \alpha(\hat{\Delta}_{nk}-\Delta^{(k)\prime}+R)\bm{x}_{\Lambda,(n',k')}$
  \EndIf
  \State $\hat{p}\gets \left(1+\exp(-\bm{\psi}_{(n',k')}^{\mathrm{T}}\bm{x}_{\Lambda,(n',k')})\right)^{-1}$
    \State $\bm{\psi}_{(n',k')}\gets \bm{\psi}_{(n',k')}-
    \alpha\left(\hat{p}-\mathbb{I}(R>0)\right)\bm{x}_{\Lambda,(n',k')}$
  \State $\Lambda=\Lambda'$
  \EndFor
\end{algorithmic}
\end{algorithm}

\section{Numerical Results}\label{sec:numres}
We consider the case with $K=20$ sources and $N=20$ sensors. To obtain a wide variety of observation probabilities, we let sensor $n$ observe updates from each source with probability $p_{nk}=1/2^n$ for $k=1,\ldots,K$. In addition to the two
algorithms presented in the previous sections (optimal and
$\epsilon$-greedy), we evaluate random sampling,
highest $\sigma_{nk}$ first, which samples the sensor-source pair with
highest $\sigma_{nk}$, and a genie-aided algorithm that is aware of
the age at each sensor and always schedules the one that results in the largest reduction in age. For $\epsilon$-greedy we use
a learning rate of $\alpha=10^{-5}$. Smaller values result in slow convergence, while larger values cause divergence due to the sporadic rewards. In order to shorten the convergence time, we perform full exploration in the first
50\,000 time slots by setting $\epsilon=1$, and then after 50\,000 time slots reduce it to $\epsilon=0.1$. Each simulation is run for a total duration of 100\,000 time slots, but to eliminate the bias of the random exploration in the $\epsilon$-greedy policy, we discard the initial 60\,000 time slots from all policies.

\begin{figure}[tb]
  \centering
  \includegraphics{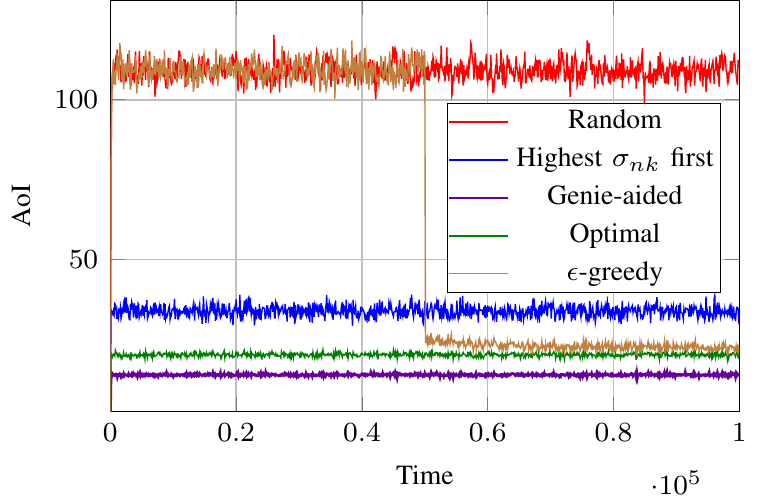}
  \caption{Average AoI for $\lambda=0.5$ and $p_{nk}=1/2^n$. The initial exploration of the $\epsilon$-greedy policy is reflected in the
    decrease in AoI after 50000 time slots.}\label{fig:timeplot}
\end{figure}

We first study the average AoI over time in a scenario where updates are generated with equal rates $\lambda=0.5$. \Cref{fig:timeplot} shows the average AoI for the different policies
averaged over 30 realizations. The optimal policy performs close to the genie-aided
policy, and the $\epsilon$-greedy algorithm also achieves a low AoI after the initial random exploration, which suggests that the system parameters can be efficiently learned and
exploited despite the simple reward model.

\begin{figure}[tb]
  \centering
  \includegraphics{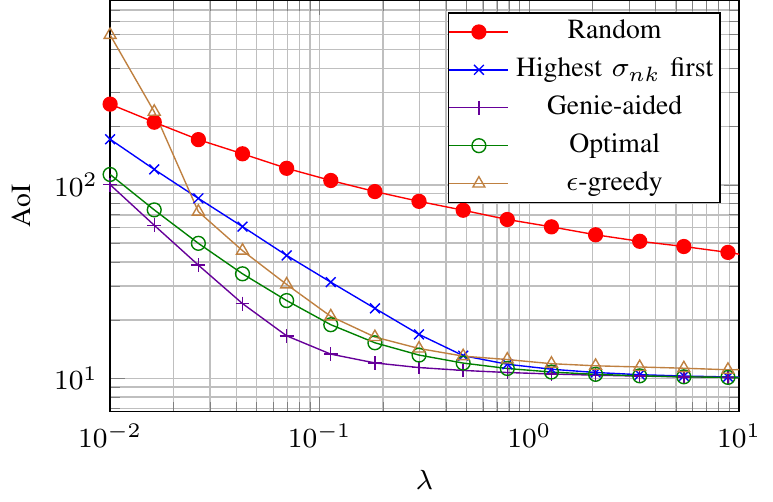}
  \caption{Average AoI with $p_{nk}=1/2^n$ for varying $\lambda$.}\label{fig:p2_vs_aoi}
\end{figure}

\Cref{fig:p2_vs_aoi} shows the same scenario for varying values of $\lambda$. As before, the optimal strategy performs close to the genie-aided strategy. Furthermore, $\epsilon$-greedy performs close to optimally for $\lambda>0.03$, while the performance is bad for low values of $\lambda$. This is due to overfitting of the prediction model when non-zero rewards occur rarely. It is likely that this can be improved by proper tuning of $\alpha$ and increasing the simulation time to allow for the slower convergence. All policies except the random approach an AoI of $K/2=10$ as $\lambda\to\infty$, which is the average age obtained if the sources are scheduled in a round-robin fashion and always provide new updates.

\section{Conclusion}\label{sec:conclusion}
We have studied sensors with common
observations using the framework of AoI. We have presented two sensor scheduling policies that can exploit the correlation and shown that they can lead to a significant reduction of the AoI compared to policies that do not take correlation into account. The first policy is optimal and
performs close to a genie-aided policy, but assumes that the
system parameters are known. The second policy, based on contextual bandits, learns the parameters by exploring the system. Although this causes a performance gap, the results indicate that the contextual bandits framework can be used to learn a good policy, but that it is sensitive to the learning rate that needs to be tuned for the specific system characteristics.

\bibliographystyle{IEEEtran}


\end{document}